\documentclass[aps,prapplied,12pt,groupedaddress,notitlepage]{revtex4-1}    
\usepackage{amsmath}
\usepackage{amssymb}
\usepackage{graphicx}
\usepackage{wrapfig}
\usepackage{setspace}
\usepackage{amsmath,amssymb,bm}

\begin{document}

\title{Temporal coupled-mode theory for thermal emission from multiple arbitrarily coupled resonators}

\author{Xin Huang}
\author{Christopher Yeung}
\author{Aaswath P. Raman}%
 \email{aaswath@ucla.edu}
\affiliation{%
Department of Materials Science and Engineering, University of California, Los Angeles, Los Angeles, CA 90095}%
\date{\today}

\begin{abstract}
Controlling the spectral response of thermal emitters has become increasingly important for a range of energy and sensing applications. Conventional approaches to achieving arbitrary spectrum selectivity in photonic systems have entailed combining multiple resonantly emissive elements together to achieve a range of spectral profiles through numerical optimization, with a universal theoretical framework lacking. Here, we develop a temporal coupled mode theory for thermal emission from multiple, arbtirarily-coupled resonators. We validate our theory against numerical simulations of complex two- and three-dimensional nanophotonic thermal emitters, highlighting the anomalous thermal emission spectra that can emerge when multiple resonators with arbitrary properties couple to each other with varying strengths.
\end{abstract}

\maketitle

\section{\label{sec:level1}Introduction}

Thermal emission is a fundamental physical process whose control is essential for a broad range of imaging, sensing and energy technologies. Conventional thermal emitters are typically incoherent, broadband, unpolarized and lack directionality. Over the last two decades, a range of photonic strategies have by contrast shown that it is indeed possible to create narrowband, polarized and/or directional thermal emitters \cite{Shchegrov00,Pralle02,Greffet02,Celanovic05,Laroche05,Lee05,Puscasu08,Shen09}.
The spectral characteristics of thermal emission in particular have been sculpted through a range of nanophotonic approaches including metallic nanoantennas \cite{Liu11,Liu17}, photonic crystals \cite{Lin00,Luo04,Lin03,Inoue14,Cornel99,Zoysa12} and semiconductor nanorods \cite{Mason11}. Tailoring the frequency and bandwidth response of emitted thermal radiation using nanophotonic strategies has in turn proved critical for improved performance in many emerging applications, including thermophotovoltaics \cite{Guo16,Lin03,Narayanaswamy03,Nagpal08,Rephaeli08,Messina13,Lenert14}, radiative cooling \cite{Raman14,Chen16,Kou17,Hossain15,Shi15,Gentle15,Goldstein17} and thermal management \cite{Hsu16,Tong15,Catrysse16,Zanjani17}. 

Conventional strategies to achieve an arbitrary degree of control over spectral selectivity for a thermal emitter rely on either numerical optimization or exploiting known electromagnetic mode behavior in conventional photonic systems. In this context, a resonance based approach to understanding thermal emission offers the potential of both understanding and designing desired thermal emission spectra in a more systematic way. Temporal coupled-mode theory \cite{Haus84,Suh04} in particular is a widely-used semi-analytical method that has been shown to provide excellent approximations and physical insight into the behavior of a range of resonance- and mode-driven nanophotonic structures and devices \cite{Khan99,H-C-Liu11,Fan03,Yanik04,Ramirez11,Ghebrebrhan11}. 

Recently, temporal coupled mode theory for single-mode thermal emitters was developed to analytically model the thermal emission from such an emitter with high accuracy \cite{Zhu13}. Further work has extended this coupled-mode theory to include coupling between multiple identical resonant thermal emitters, and shown that thermal emission decreases with increasing number of identical thermal emitters in analogy to quantum superradiance effects \cite{Zhou15}. As a semi-analytical model, coupled-mode theory is able to offer deeper insight on the effect of material and structural parameters and the resulting spectral nature of their thermal emission, an insight that simulations alone cannot offer. However, prior work has been limited to describing a constrained set of thermal emitters due to the resonators being identical in character. For many practical scenarios, highly complex and selective thermal emission spectra with multiple frequency peaks and varying bandwidths are desirable, but difficult to achieve with single-mode resonators, and more generally are challenging to elucidate. One particularly promising mechanism to achieving complex thermal emission spectra is to consider systems supporting multiple arbitrary resonators \cite{Liu11,Liu17} which may additionally couple or hybridize with each other, thereby enhancing or suppressing thermal emission with several degrees of potential freedom to control the resulting spectral response. Such a scenario, while intriguing, is however challenging to model analytically and can require many complex and slow simulations. A universal theoretical framework for understanding the full range of complex coupling, and resulting thermal emission that may be possible in multi-resonant photonic structures is currently lacking, but could enable deeper insights into the behavior of such structures and enable more rapid design of their arrangements.

In this Letter we introduce an extended temporal coupled-mode theory framework to derive an analytical formalism to model thermal emission from an arbitrary number of resonators which  can arbitrarily couple to each other. We validate the theory against simulation-based calculations of thermal emission from a range of physically realizable systems. We first demonstrate the coupled-mode theory's accuracy for multiple two-dimensional slit resonators, where the dielectric permittivity of each slit can vary, along with the distance between each slit, thereby altering both the real and imaginary part of the coupling coefficient. We also demonstrate the coupled-mode theory's capabilities with a three slit system, highlighting how both subradiant enhancement and superradiant suppression of thermal emission can be selectively engineered at different frequencies. Finally, we demonstrate the accuracy of the coupled-mode theory in predicting the thermal emission spectra of supercell three-dimensional metal-dielectric-metal resonators for a range of inter-resonator distances. Collectively, our results offer a general theoretical framework capable of taking the response of individual resonators and using them to determine their complex spectral response when integrated and hybridized with each other.

\section{Extended Coupled-mode Theory}
We first develop an extended temporal coupled-mode theory capable of describing the collective thermal emission from $N$ different emitters that are coupled to each other with varying degrees of strength. Each resonant emitter is assumed to have an amplitude $\textbf{a}=(a_1,a_2,...,a_N)^T$. The energy stored inside an emitter may decay through three pathways. The first pathway is through intrinsic absorption which is represented by the intrinsic decay rate $\bm{\Gamma_0}$=diag($\gamma_{01}$, $\gamma_{02}$ ...$\gamma_{0N}$). The second is to decay to the external free space channels, which is described by the external decay rate $\bm{\Gamma_e}$ =diag($\gamma_{e1}$, $\gamma_{e2}$ ...$\gamma_{eN}$). The third is through coupling with other resonators, which is expressed by a complex coupling coefficient matrix $K$ defined as
\begin{equation}
\mathbf{K} = \left ( \begin{array}{ccccc}
0 & \kappa_{12}+i\beta_{12} & \cdots & \kappa_{1(N-1)}+i\beta_{1(N-1)} & \kappa_{1N}+i\beta_{1N} \\
\kappa_{21}+i\beta_{21} & 0 & \cdots & \kappa_{2(N-1)}+i\beta_{2(N-1)} & \kappa_{2N}+i\beta_{2N}\\
\vdots & & \ddots & & \vdots \\
\kappa_{(N-1)1}+i\beta_{(N-1)1} & \kappa_{(N-1)2}+i\beta_{(N-1)2} & \cdots & 0 & \kappa_{(N-1)N}+i\beta_{(N-1)N}\\
\kappa_{N1}+i\beta_{N1} & \kappa_{N2}+i\beta_{N2} & \cdots & \kappa_{N(N-1)}+i\beta_{N(N-1)} & 0 \end{array} \right )
\end{equation}.
To capture arbitrary coupling between resonating elements, we define $\kappa$ as the real part of the coupling coefficient, which in our context can capture scenarios such as variable distance between resonators, and $\beta$ which is the imaginary part of the coupling strength, and captures the phase difference between resonators. In the framework of the fluctuation-dissipation theorem, the absorption process is balanced by a random thermal excitation source $\bm{n}$. With all these considerations, the dynamic equations for resonance amplitudes can be written in the following form:
\begin{eqnarray}
 \frac{{d\bm{a}}}{dt} =(j\bm{\Omega_0}-{\bm{\Gamma_0}-\bm{\Gamma_e}}) \ \bm{a} +\sqrt{2\ \bm{\Gamma_0}} \ \bm{n}-\bm{K a}
\end{eqnarray}
where $\bm{\Omega_0}$ = diag($\omega_1$, $\omega_2$ ... $\omega_N$) describes the resonant frequency of each resonator. We can explicitly calculate $\bm{a(\omega)}$ in the frequency domain, 
\begin{eqnarray}
 \bm{a(\omega)} =(j(\omega-\bm{\Omega_0})+{\bm{\Gamma_0}+\bm{\Gamma_e}}+\bm{K})^{-1} \sqrt{2\ \bm{\Gamma_0}} \ \bm{n}
\end{eqnarray}
We normalize the amplitude so that the mode energy is given by $|\bm{a}|^2$. Here we have introduced a noise source vector $\bm{n}$ in Eq. (1), to compensate for the intrinsic resonator loss and maintain thermal equilibrium \cite{Zhu13}. Following the fluctuation-dissipation theorem this noise source is defined by a correlation function (see Supplementary Information):
 \begin{eqnarray}
 \langle\bm{n}^{*}(\omega) \bm{n}(\omega')\rangle=\frac{1}{2N\pi} \bm{\Theta}(\omega,T) \delta(\omega-\omega')
 \end{eqnarray}
where $\bm\Theta(\omega,T)=\frac{\hbar \omega}{e^\frac{\hbar \omega}{kT}-1}$. The total power emitted as thermal radiation, $\langle\pmb{P}\rangle$, can then be calculated as 
\begin{eqnarray}
\langle\bm{P}(t)\rangle&=&2\bm{\Gamma_e}\langle\bm{a}^{*}(t)\bm{a}(t)\rangle  \nonumber \\
&=&2\bm{\Gamma_e}\int_{0}^{\infty}d\omega\int_{0}^{\infty}d\omega'e ^{-j(\omega-\omega')t}\langle\pmb{a}^{*}(\omega)\pmb{a}(\omega')\rangle \nonumber\\    
&=&\int_{0}^{\infty}d\omega\frac{\bm{\Theta}(\omega,\bm{T})}{2N\pi}4\bm{\Gamma_e}((j(\omega-\bm{\Omega_0})+\bm{\Gamma_0}+\bm{\Gamma_e}+\bm{K})^{-1})^2 \bm{\Gamma_0}
\end{eqnarray}
Solving for the power spectral density of thermal emission we find that
\begin{eqnarray}
\bm{P}(\omega)=\frac{\bm{\Theta}(\omega,\bm{T})}{2N\pi}4\bm{\Gamma_e}((j(\omega-\bm{\Omega_0})+\bm{\Gamma_0}+\bm{\Gamma_e}+\bm{K})^{-1})^2 \ \bm{\Gamma_0}  
\end{eqnarray}
\noindent Here, as in conventional expositions of temporal coupled mode theory for multi-port systems, we assume that each resonances can decay into every other port in the system, described by a total value $d_{ij}$ and encompassed by a coupling matrix $\mathbf{D}$:

\begin{equation}
\mathbf{D} = \left ( \begin{array}{ccccc}
d_{11} & d_{12} & \cdots & d_{1(N-1)}& d_{1N} \\
d_{21} & d_{22} & \cdots & d_{2(N-1)} & d_{2N}\\
\vdots & & \ddots & & \vdots \\
d_{(N-1)1} & d_{(N-1)2} & \cdots & d_{(N-1)(N-1)} & d_{(N-1)N}\\
d_{N1} & d_{N2} & \cdots & d_{N(N-1)} & d_{NN} \end{array} \right ).
\end{equation}.

\noindent By energy conservation the total coupling matrix $\mathbf{D}$ bounds the values of the coupling coefficients to each relevant channel (see Supplementary Information):
\begin{eqnarray}
2(\bm{\Gamma_0}+\bm{\Gamma_e}+\bm{K}) = \bm{D^+D}\
\end{eqnarray}
Eq. (6) is a key result of this paper and provides a general expression for thermal emission from $N$ resonators that are coupled to each other arbitrarily. To elucidate the power of this result, we first write analytical forms for small $N$ scenarios of typical interest, beginning with the $N$ = 1 scenario where coupling is not relevant:
\begin{eqnarray}
P(\omega)=\frac{\bm{\Theta}(\omega,\bm{T})}{2\pi}\frac{4\gamma_{01} \gamma_{e1}}{(\omega-\omega_1)^2+(\gamma_{01}+\gamma_{e1})^2}
\end{eqnarray}
This expression is simply a standard Lorentzian form of power emitted due to a single resonance previously derived in Ref. \cite{Zhu13}. For two resonators ($N$ = 2), however, we must include  complex coupling terms $\kappa_{ij}+i\beta_{ij}$ for generality. This results in the following expression for the radiated power:
\begin{widetext}
\begin{eqnarray}
P(\omega)&=&\frac{\bm{\Theta}(\omega,\bm{T})}{4\pi}(\frac{4\gamma_{01} \gamma_{e1}\alpha_2^2+4\gamma_{e1}\gamma_{02}\kappa_{12}^2+4\gamma_{e1}\gamma_{02}\beta_{12}^2-8\alpha_2\kappa_{12}\gamma_{e1}\sqrt{\gamma_{01}\gamma_{02}}}{(\alpha_1\alpha_2-(\kappa_{12}+i\beta_{12})(\kappa_{21}+i\beta_{21}))^2} \nonumber\\
&&+\frac{4\gamma_{e2}\gamma_{02}\alpha_1^2+4\gamma_{e2}\gamma_{01}\kappa_{21}^2+4\gamma_{e2}\gamma_{01}\beta_{21}^2-8\alpha_1\kappa_{21}\gamma_{e2}\sqrt{\gamma_{01}\gamma_{02}}}{(\alpha_1\alpha_2-(\kappa_{12}+i\beta_{12})(\kappa_{21}+i\beta_{21}))^2})
\end{eqnarray}
\end{widetext}
Here for convenience we have introduced variables $\alpha_j$ defined as:
$$
\alpha_j=i(\omega-\omega_j)+\gamma_{0j}+\gamma_{ej} 
$$
$\kappa_{12}$ and $\kappa_{21}$ are the real parts of the coupling term determined by the spatial distance between the resonances, while the complex terms $\beta_{12}$ and $\beta_{21}$ describe the phase mismatch between resonances. Due to energy conservation, a two resonator system is constrained by Eq. (8), resulting in the following expression:
\begin{equation}
\cos(\theta_{12}-\theta_{11}) + i \sin(\theta_{12}-\theta_{11})+  
\cos(\theta_{22}-\theta_{21})+i \sin(\theta_{22}-\theta_{21})
=\frac{2\kappa_{12}+2j\beta_{12}}{\sqrt{(\gamma_{01}+\gamma_{e1})(\gamma_{02}+\gamma_{e2})}}
\end{equation}
Here, $\theta_{ij}$ is the phase angle of $d_{ij}$. As in conventional coupled-mode theory, this energy conservation relation extends to arbitrary numbers of resonators and fundamentally links the various coupling terms external to each resonator: coupling to neighboring resonators or to free space. In the two resonator case, we can first consider the scenario when the two resonators are spatially closed each other. In this case, as the two resonators move closer and closer, $\theta_{12}$ will be getting closer to $\theta_{11}$ and $\theta_{21}$ will be getting closer to $\theta_{22}$. Therefore, cos($\theta_{12}-\theta_{11}$) is near 1 and sin($\theta_{12}-\theta_{11}$) is near 0, and the total thermal emission peak power is strongly driven by the real part of the coupling coefficients. When the two resonators are physically far away from each other, as $\theta_{12}$ and $\theta_{21}$ is small, the coupling strength is determined by the phase angle of $\theta_{11}$ and $\theta_{22}$. If $\theta_{11}$ and $\theta_{22}$ closes to $\pi/2$, the total thermal emission peak is mainly determined by the imaginary portion of the coupling strength which encodes phase mismatch. If $\theta_{11}$ and $\theta_{22}$ closes to 0, the total thermal emission peak power is strongly driven by the real part of the coupling coefficients. More generally, the behavior of each resonant peak in an arbitrary multiple-resonance system is determined by both real and imaginary parts of the coupling coefficients.

\section{Numerical Results}

To validate the extended couple-mode theory developed above, we consider an exemplary system consisting of multiple narrow dielectric slits of arbitrary permittivity which are introduced into a perfect electric conductor (PEC) layer, as shown in Fig. 1(a). In the system, the optical fields are confined in the slit and can couple with each other. This system is an extension of the slit resonator system considered in Ref. \cite{Zhou15} which consider multiple identical resonators that were sufficiently close to each other to enable near-field coupling. 
\begin{figure}[t]
\centering
\includegraphics[width = 1.0\textwidth]{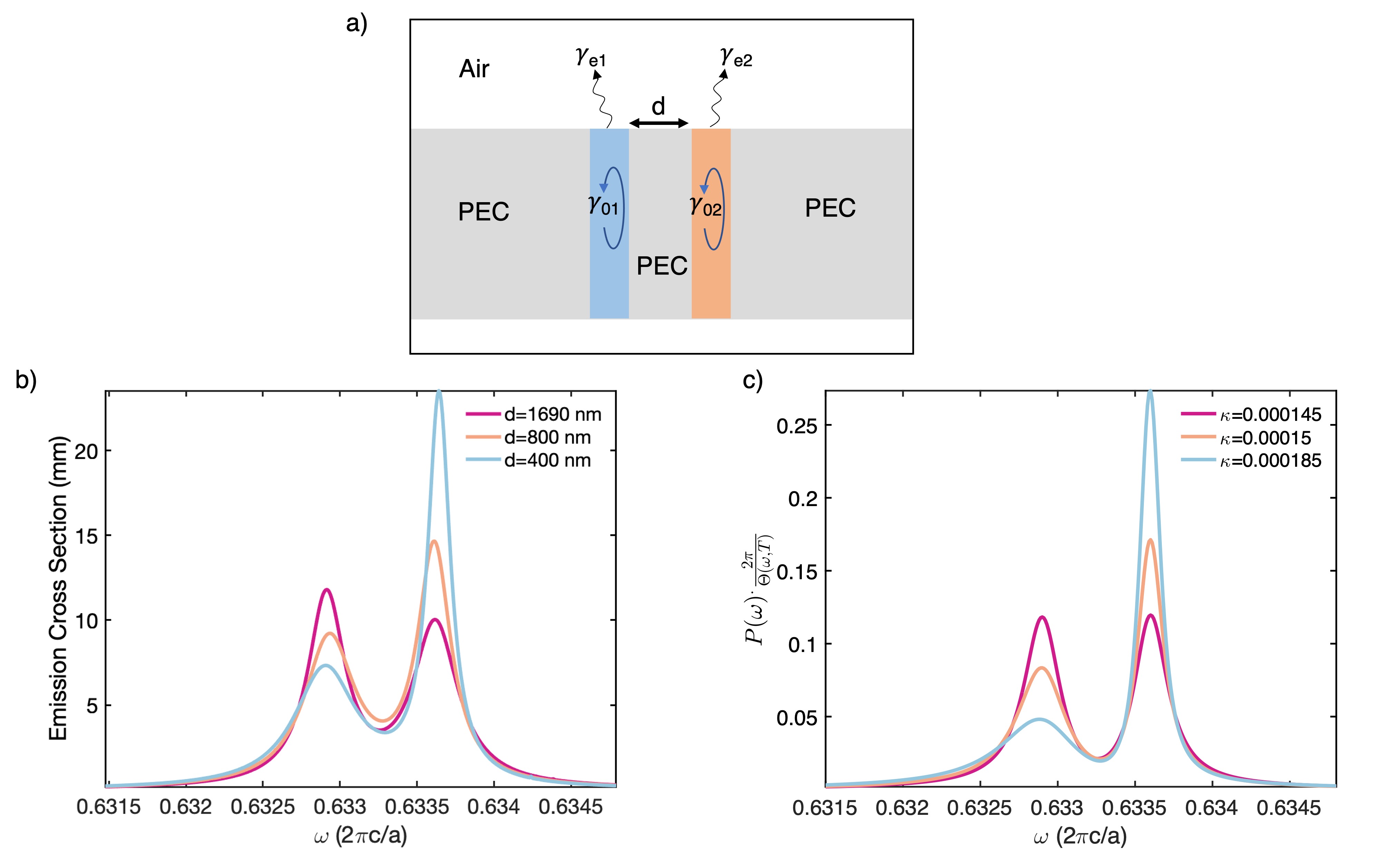}
\caption{\textbf{The effect of varying distance between dissimilar resonators.} a) The structure of resonant emitters consisting of two slits in two PEC slabs. The length and width of the slit are 1.4 $\mu$m and 5 nm, respectively. They are filled with two different emissive materials with a dielectric constant of $\varepsilon_1$ = 12.5 + 0.001$i$ and $\varepsilon_2$ = 12.53 + 0.001$i$. b) FDFD-simulated spectra of emission cross section for two resonant emitters with a distance of 400 nm, 800 nm and 1690 nm, separately. c) Coupled mode theory thermal emission power prediction for two different emitters with a coupling strength of $\kappa$ = 0.000145, $\kappa$ = 0.00015, $\kappa$ = 0.000185 }
\end{figure}

We first consider the two resonator (two dielectric slit) scenario and demonstrate how the coupled-mode theory accurately predicts the effect of varying distance between dissimilar resonators, as well as differing permittivities at a fixed separation distance. In this scenario, the slits are 1.4 $\mu$m wide, and 5 nm long, and contain dielectric media with permittivities $\varepsilon_1 = 12.5 + 0.001i$ and $\varepsilon_2 = 12.53 + 0.001i$ respectively. Since the permittivity of each slit is different the system supports resonances at slightly different frequencies, $\omega_1$ and $\omega_2$ as can be seen in Fig. 1(b). We then calculate the emission cross section of this system using the finite-difference frequency domain method for different values of the slit separation distance $d$ as shown in Fig. 1(b). Simulations reveal a notable behavior as we decrease the distance between two resonators: the resonator with lower dielectric permittivity sees a large enhancement in its total associated emission while a large decrease for the resonator with higher permittivity is observed. 

We next model the same system using the extended coupled-mode theory and compare its predictions to the FDFD simulations. By defining a purely real inter-resonator coupling constant, $\kappa$ and examining a range of its values we are able to systematically replicate the trend observed in Fig. 1(b) in Fig. 1(c) remarkably well. In particular, as the two resonators are moved closer, the real part of the coupling strength further increases. This in turn influences the internal coupling coefficients $\gamma_{01}$ and $\gamma_{02}$ and external coupling coefficients $\gamma_{e1}$ and $\gamma_{e2}$ due to energy conservation. As detailed in the Supplementary Information, further increasing the real part of coupling constant will need to increase $\sqrt{(\gamma_{01}+\gamma_{e1})(\gamma_{02}+\gamma_{e2})}$ thereby resulting in a decrease of the lower frequency peak, and an increase of the higher frequency peak. The exact same behavior is observed in the FDFD simulations, with a small variation in resonance frequency the only notable difference. This result suggests that once coupling coefficients have been established for a particular multi-resonator system of interest, the temporal coupled-mode theory can be used as a rapid simulator of thermal emission from the class of nanophotonic structures being examined.

\begin{figure}[h]
\centering
\includegraphics[width = 1.0\textwidth]{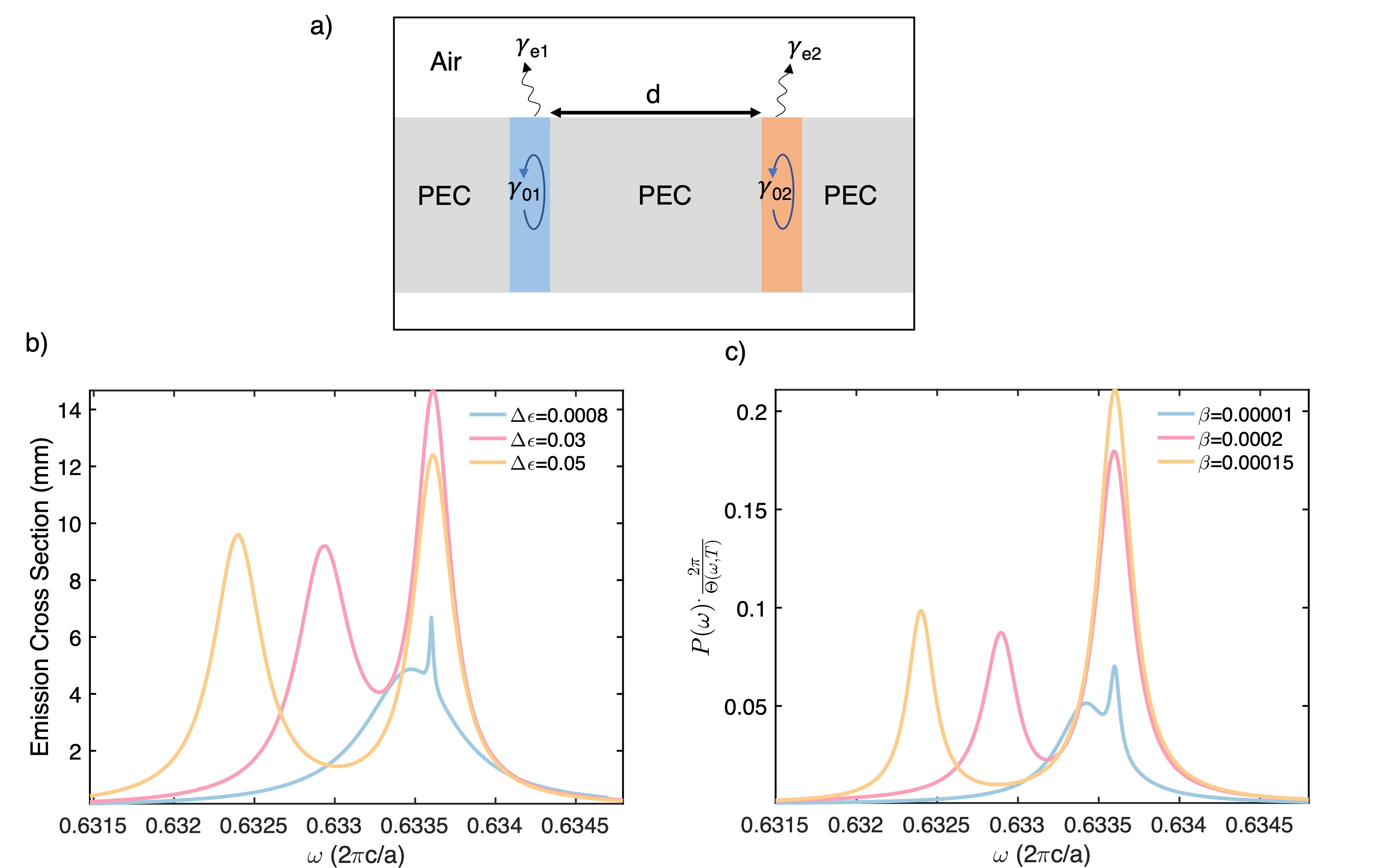}
\caption{\textbf{The effect of changing permittivity in one of the two dielectric slits at a fixed, large distance.} a) The structure of resonant emitters consisting of two slits in two PEC slabs. The length and width of the slit are 1.4 $\mu$m and 5 nm, respectively. They are filled with two different dielectric materials with non-zero emissivity that are separated by a constant distance $d$ = 800 nm.  b) FDFD-simulated spectra of emission cross section for two resonant emitters. The dielectric constant difference of $\Delta \epsilon_1$ = 0.0008, $\Delta \epsilon_2$ = 0.03 and $\Delta \epsilon_3$ = 0.05 separately. c) Coupled mode theory thermal emission power prediction for two different emitters with a different imaginary coupling strength of $\beta$=0.00001, $\beta$ = 0.0002, $\beta$ = 0.00015.}
\end{figure}

Next, we examine the effect of changing the permittivity in one of the two dielectric slits at a fixed, large distance $d$ = 800 nm (Fig. 2(a)). As the permittivity of one of the slits is changed, the imaginary part of inter-resonator coupling coefficient, $\beta$ changes due to phase difference between each resonator. Since the resonators are far apart, we expect minimal contribution from the real part of the coupling coefficient $\kappa$. In our system, we maintain the permittivity of $\varepsilon_1$ as 12.5 + 0.001$i$ and change the permittivity of $\varepsilon_2$ to 12.5008 + 0.001$i$, 12.53 + 0.001$i$, and 12.55 + 0.001$i$ respectively. FDFD simulations shown in Fig. 2(b) highlight that as $\varepsilon_2$ changes the emission peak associated with resonator 2 increases first, reaching a peak at $\Delta \varepsilon = 0.03$ but then decreasing as $\varepsilon_2 $ is further increased. We explore the same system using coupled-mode theory in Fig. 2(c) and find a range of $\beta$ values which result in exactly the same behavior observed in the FDFD simulations, further demonstrating the utility of this coupled-mode theory model. Indeed, we emphasize that a purely real coupling coefficient would be insufficient to capture the scenarios shown here. Referring back to Eq. (11) we observe that as the sin($\theta$) term purely influences $\beta$ it thereby encodes the phase difference between the resonators, which peaks at $\theta = \pi/2$ which corresponds to a particular $\beta$ value depending on the strengths of the internal coupling coefficients $\gamma_0$ and external coupling coefficients of the resonances $\gamma_e$.

To highlight the generality and flexibility of the developed coupled-mode theory we now examine a more complex systems involving three dielectric slits separated by some arbitrary distance in a finite thickness PEC slab, and filled with dielectric materials of arbitrary permittivity, as is shown in Fig. 3(a). We consider the scenario where the first and third resonators have the same permittivity $\varepsilon_1 = \varepsilon_3 =$ 12.5 + 0.001$i$ while the middle resonator has a slightly different permittivity of $\varepsilon_2 =$ 12.53 + 0.001$i$. In Fig. 3(b), we compare the emission cross section as simulated by FDFD for this three slit resonator systems against a two resonator only system with $\varepsilon_1 =$ 12.5 + 0.001$i$ and $\varepsilon_2 =$ 12.53 + 0.001$i$ and without $\varepsilon_3$ . Remarkably, we observe that the addition of the third resonator suppresses the lower frequency peak (associated with $\varepsilon_1$ and $\varepsilon_3$) while greatly enhancing the emission peak associated with the middle resonator (with permittivity $\varepsilon_2$). Coupling between the resonators 1 and 3 results in a super-radiant suppression of emission analogous to that observed in Ref. \cite{Zhou15} for the lower frequency peak. However, coupling between resonator 2 and its neighbors results in a dramatic sub-radiant enhancement in its thermal emission peak. We can model both the baseline two resonator, and the more complex three resonator system using the coupled-mode theory and find that it replicates the result observed in the full-wave electromagnetic simulation, as is shown in Fig. 3(c). Our results, in addition to highlighting the capabilities of the coupled-mode theory, show a new mechanism to develop high power, narrow bandwidth thermal emitters through inter-resonator coupling.

\begin{figure}[h]
\centering
\includegraphics[width = 1\textwidth]{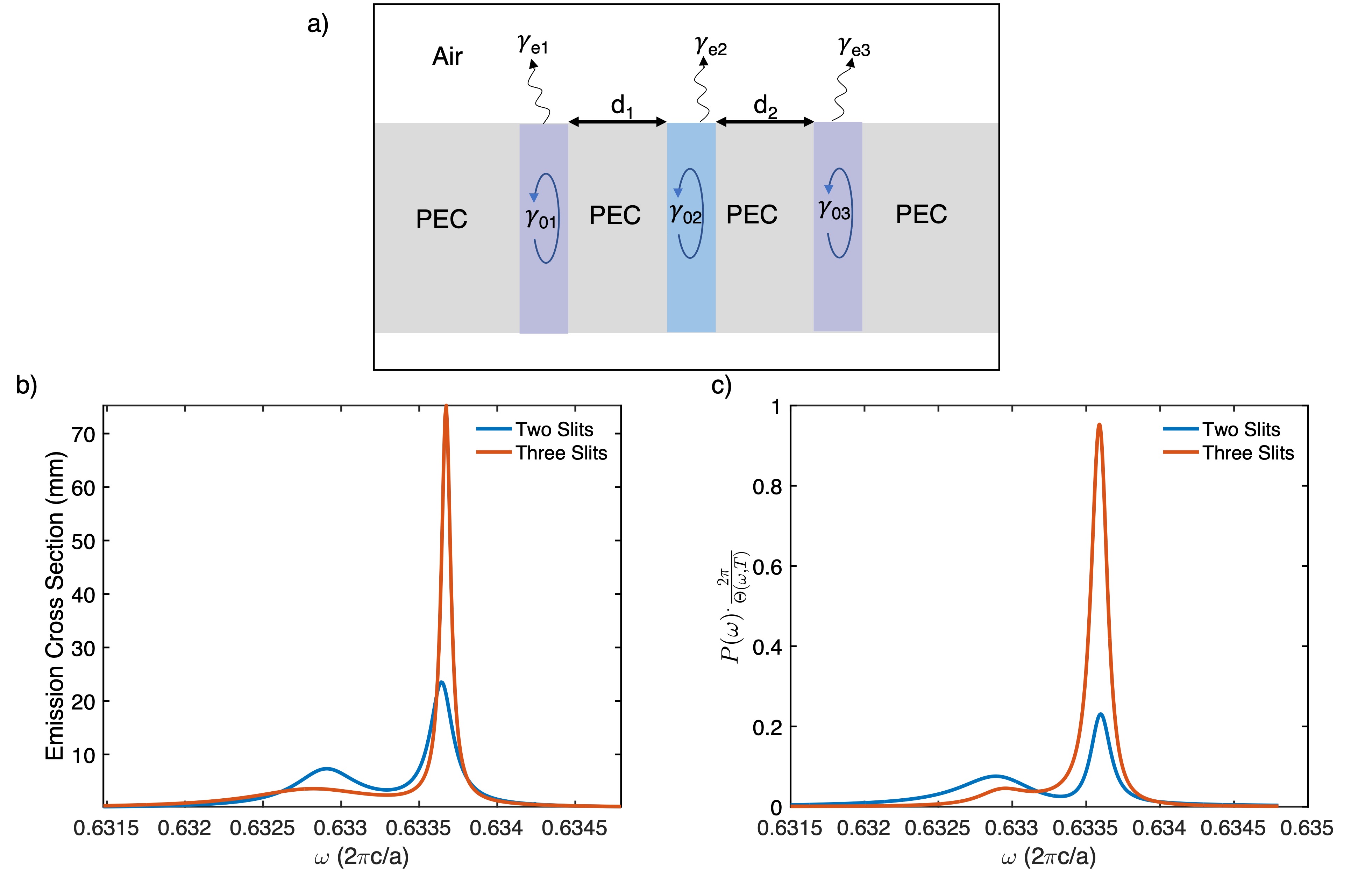}
\caption{ \textbf{Comparison between two resonant emitters system and three resonant emitters system.} a) The structure of resonant emitters consisting of three slits in three PEC slabs. The length and width of the slit are 1.4 $\mu$m and 5 nm, respectively. They are filled with three emissive materials. b) FDFD-simulated spectra of emission cross section for two resonators with a permittivity $\varepsilon_1$ = 12.5 + 0.001i and $\varepsilon_2$ = 12.53+0.001i, and for a three resonant emitter scenario where the dielectric permittivities are $\varepsilon_1$ = 12.5 + 0.001$i$, $\varepsilon_2$ = 12.53 + 0.001$i$ and $\varepsilon_3$ = 12.5 + 0.001$i$. c) Coupled mode theory thermal emission power prediction for both the two resonator and three resonator scenarios, showing strong alignment with the numerical simulations of b).}
\end{figure}

Finally, we demonstrate the ability of the coupled-mode theory to accurately model thermal emission from complex, three-dimensional nanophotonic structures. In particular we consider complex supercells of metal-insulator-metal (MIM) metasurfaces \cite{Liu11,Chris21} featuring cross-shaped resonators made of two different noble metals, schematically shown in Fig. 4(a). The structures have $\pi$/2 rotational symmetry and consist of three layers. The top layer consists of a pair of gold metasurface elements, and a pair of silver metasurface elements with arms lengths of $l_1=2.5\ \mu$m, $l_2=2.1\ \mu$m, and arm widths $w_1 = 0.6\ \mu$m , $w_2 = 0.4\ \mu$m respectively, with a thickness of 0.1\ $\mu$m. The permittivities of both silver and gold are modeled with a Drude fit that is accurate to the long-wave infrared target wavelength range \cite{Ordal83}. The metasurfaces lie atop a 0.2 $\mu$m Al$_2$O$_3$ layer whose permittivity is defined in our wavelength region of interest with a Drude-Lorentz model $\varepsilon=\varepsilon_\infty-\sigma/(\omega^2-i\omega\gamma-\omega_0^2)$, with $\varepsilon_\infty$ = 2.228, $\sigma$ = 0.008385$(2\pi c/a)$, $\gamma$ = 0.04$(2\pi c/a)$, and $\omega_0$ = $0.08(2\pi c/a)$, where $a=1\ \mu m$. The bottom layer is assumed to be gold as well at a thickness of 0.2 $\mu m$. 
\begin{figure}[h]
\centering
\includegraphics[width = 1.0\textwidth]{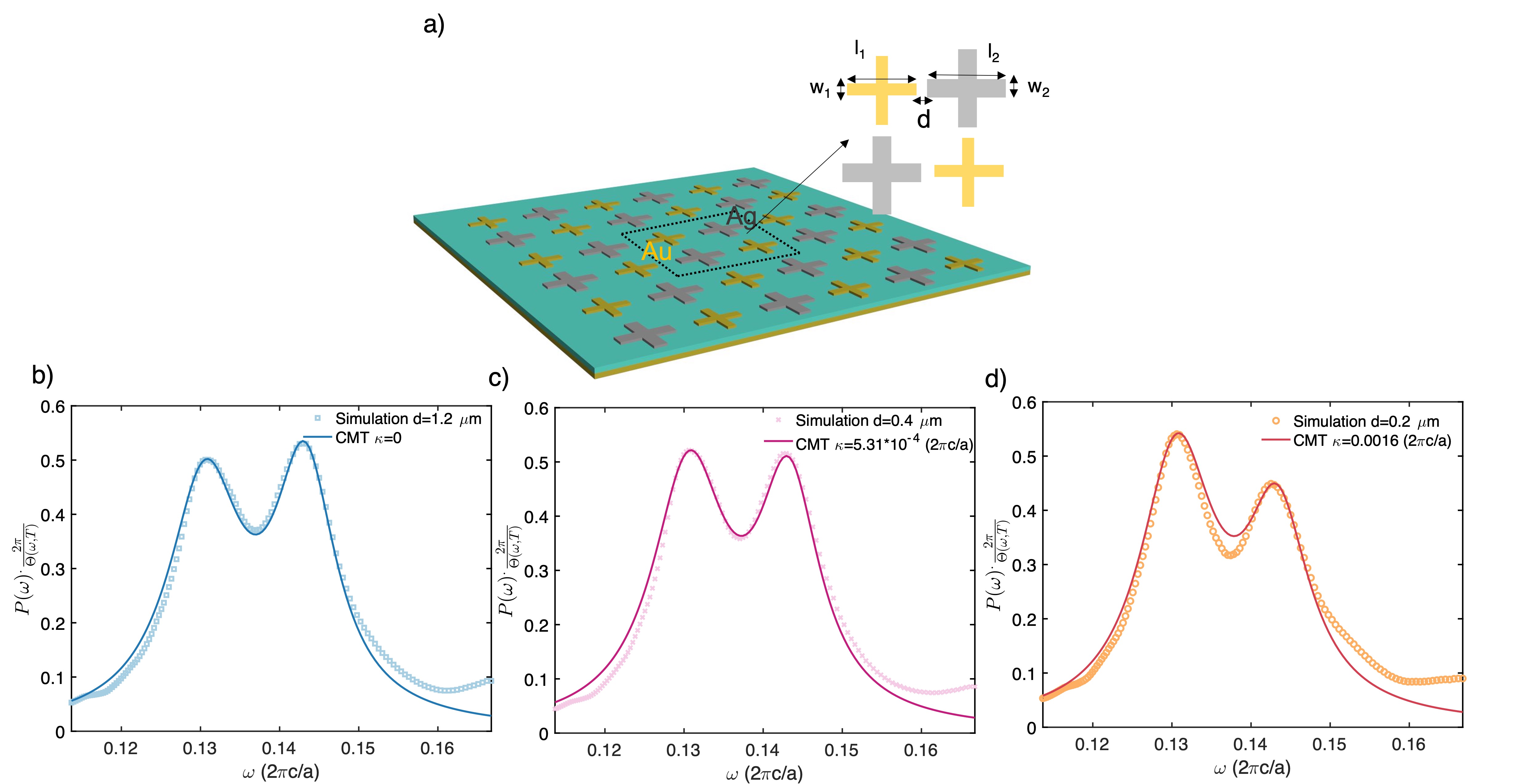}
\caption{\textbf{Couple-mode theory predictions vs. full-field simulations of a complex supercell metasurface.} a) Geometry of the supercell metasurface architecture evaluated. The metasurface elements have dimensions of $l_1=2.5\ \mu$m, $l_2=2.1\ \mu$m, and $w_1 = 0.6\ \mu$m , $w_2 = 0.4\ \mu$m respectively, with a layer thickness of 0.1 $\mu$m. The central layer is Al$_2$O$_3$ with thickness 0.2 $\mu m$ while the bottom layer is gold with thickness of 0.2 $\mu$m. b) Simulation thermal emission power when d=1.2 $\mu$m versus analytical coupled-mode theory prediction when $\kappa$=0. c) Simulation thermal emission power when d=0.4 $\mu$m versus analytical coupled-mode theory prediction when $\kappa$=5.31$\times$ $10^{-4}$(2$\pi$c/a) d) Simulation thermal emission power when d=0.2 $\mu$m versus analytical coupled-mode theory prediction when $\kappa$=0.0016 (2$\pi$c/a)}
\end{figure}
We performed full-field electromagnetic simulations of a range of MIM supercell metasurfaces where the distance between the resonators in each supercell is modified, in analogy to the scenario explored in Fig. 1 in a two-dimensional systems. We simulated each geometry across both polarizations and all angles of incidence to calculate the total spectral hemispherical emittance. Given the four-fold symmetry of the supercells, to find the total emitted power we integrate the incident angle $\theta$ from 0 to $\pi$/2 at $\pi$/12 step and the azimuth angle $\phi$ from 0 to $\pi$/4 with $\pi$/12 steps:
\begin{eqnarray}
\varepsilon (\lambda)=\frac{\int_{0}^{2\pi}\int_{0}^{\pi/2}\varepsilon(\lambda,\theta,\phi)\cos\theta \sin\theta ~d\theta d\phi}{\int_{0}^{2\pi}\int_{0}^{\pi/2}\cos\theta \sin\theta ~d\theta d\phi} 
\end{eqnarray} 

As shown in Fig. 4(b-d), the simulations show that the supercell structure supports two non-degenerate modes at $\omega_1$=0.131 (2$\pi$c/a) and $\omega_2$=0.143 (2$\pi$c/a). We then use the coupled-mode theory to fit each resonator’s intrinsic decay rate and external decay rate by simulating its hemispherical emittance with a single resonator model, finding $\gamma_{01}$ = 0.0019 (2$\pi$c/a), $\gamma_{02}$ = 0.0017 (2$\pi$c/a), $\gamma_{e1}$ = 0.0038 (2$\pi$c/a), and  $\gamma_{e2}$= 0.0030 (2$\pi$c/a). As the distance $d$ is varied, the numerically simulated total hemispherical emittance is then compared against the coupled-mode theory predictions and shows remarkably strong agreement, as can be seen in Fig. 4(b-d) for a range of resonator distances. Our results highlight a powerful capability enabled by the coupled-mode theory framework in the context of thermal emission from such complex supercell systems: simulating the response of each metasurface resonator in a periodic configuration provides sufficient information to rapidly model the behavior of complex arrangements of metasurface resonators through the use of the complex coupling coefficients $\bm{K}$.

\section{\label{sec:level1}Concluding Remarks}
In conclusion, we have developed a coupled mode theory that accurately models thermal emission from complex photonic structures composed of multiple resonators that can be arbitrarily coupled to each other. We demonstrated that the coupled-mode theory accurately models anomalous, complex behavior such as combined sub- and super-radiant thermal emission in a single nanophotonic structure, as well as the response of supercell three-dimensional MIM metasurface thermal emitters. Furthermore, the coupled mode theory provides accurate predictions while taking far less time than full-field simulations which takes hours to simulate\cite{Chris21}, and in the process provides physical insight into the original of the complex spectral behavior that can result. Although we only fit our theoretical framework using regular nanophotonic systems, this theory also implies that, once intrinsic decay rate, external decay rate and coupling coefficients for a particular multi-resonator system of interest have been determined, the temporal coupled-mode theory can be utilized as a quick simulator of thermal emission from the class of complex nanophotonic structures under consideration. As a semi-analytical model, it can provide deeper insight into the effect of material and structural characteristics on the spectral nature of their thermal emission, which simulations alone cannot provide. One can thus imagine building libraries of individual resonator responses which can then rapidly be assessed for their integration into coupled arrangements based on the developed coupled-mode theory. As demands on the complexity of the spectral response of thermal photonic emitters grow, such a framework may prove critical to further enhance and rapidly model their capabilities. 

\section{Acknowledgments}
This material is based upon work supported by the National Science Foundation (NSF CAREER) under Grant No. 2146577, and the Sloan Research Fellowship (Alfred P. Sloan Foundation).
\bibliography{Reference}
\end{document}